\documentclass{PoS}

\title{System and event activity dependent inclusive jet production with ALICE }

\ShortTitle{Jet production in pp and Pb-Pb collisions}
\author{\speaker{Yongzhen Hou}\thanks{Supported by the National Science Foundation of China (Grant No. 11505072 and 11975005) and the Fundamental Research Funds for the Central Universities (CCNU19QN060).}~ for the ALICE Collaboration  \\
        Key Laboratory of Quark \&{} Lepton Physics of Ministry of Education, Institute of Particle Phisics, Central China Normal University \\
        E-mail: \email{yongzhen.hou@cern.ch}}

\abstract{
Jets are produced by processes involving high momentum transfer 
of initial partons at high energies.
Comparing jet production in pp and nucleus-nucleus collisions will allow us to study
the jet-quenching effect caused by the hot and dense QCD medium produced in nucleus-nucleus collisions when energetic partons traverse the medium. In particular, systematic studies of jet production in different multiplicity environments will provide in-depth understanding of the medium properties and their evolution from small to large systems. In small systems and high multiplicity events, 
the bulk properties extracted by the low transverse momentum particle production behaves as if a hot QCD medium was created, but such behaviour is not observed with hard probes. Study of jet production in different multiplicity proton-proton collisions then helps to explore the QGP existence in small systems.

In this proceeding, the jet cross section measurements in different collision systems using the data taken by ALICE during the LHC Run 2 are presented. The nuclear modification factor of jets are presented to characterize the jet-quenching effect.  We observe that more jets are produced in high multiplicity bins compared to the inclusive one, while the jet production enhancement in high multiplicity environments has weaker jet $p_{\rm T}$ or resolution parameter dependences. In order to study the jet collimation properties, the jet cross section ratios for different jet resolution parameters are also measured and compared to theoretical models. As expected, the jets get more collimated at high $p_{\rm T}$ in numerous multiplicity bins, with no collision energy or multiplicity dependence when compared to earlier results. 
}

\FullConference{
European Physical Society Conference on High Energy Physics - EPS-HEP2019 -\\
			10-17 July, 2019\\
			Ghent, Belgium}

\begin{document}

\section{Introduction}
A jet is spray of hadrons from high momentum hard scattered partons fragmentation which collimate in finite space.
The jet cross section measurements set constraints on Perturbative Quantum Chromo-Dynamics (pQCD) calculations and provide significant tools for studying Standard Model and Beyond Standard Model physics~\cite{ref:pp7tev}. By using different jet resolution parameters, the splitting function of partons in vacuum can also be investigated, which is close to the original parton collimation behaviours.

In high-energy nuclear collisions, jet production suppression or jet quenching has been proposed as a good probe of the hot and dense medium production in relativistic heavy-ion collisions. The quenching effect of energetic partons is caused by multiple scattering and induced parton energy loss during their traverse of hot QCD medium~\cite{ref:jetquenching}. The inclusive jet production measurements in pp collisions thus provide a vital reference for similar measurements in nucleus-nucleus (A-A), proton-nucleus (p-A) and high multiplicity proton-proton (pp) collisions~\cite{ref:inclusivemult}.  
 
So far, a large number of experimental observables have been studied to probe the QGP properties. A QGP medium has been expected to be created in the heavy-ion collisions. In addition, by studying the QGP observables in high multiplicity proton-proton collisions, we have observed similar behaviour as shown in heavy-ion collisions where QGP exists~\cite{ref:multstrange}. Whether QGP can be formed in high multiplicity  collisions of small systems, such as pp and p-Pb, is an open question that requires further investigation.
Therefore, a detailed study of the jet production in different collision systems and event activities with various jet resolution parameters will help investigate further the existence of medium effects on jets in small systems with high multiplicity.

Using the large volume of LHC Run 2 data taken by ALICE detector~\cite{ref:ALICE}, we will first present the jet production yields in Pb-Pb collision at the center-of-mass energy of $\sqrt{s_{\rm NN}}$ = 5.02 TeV plus nuclear modification factors to study the jet quenching effect. We have also analyzed the jet production yields in proton-proton collisions at the center-of-mass energy of $\sqrt{s}$ = 13 TeV, and obtained the charged jet cross sections in different multiplicity intervals. The jet cross sections are also measured using different jet resolution parameters to study the jet collimation behaviour. 

\section{Experimental setup and analysis strategy}
ALICE is a dedicated heavy-ion experiment at LHC at CERN, and its performance is described in~\cite{ref:ALICE}. 
We report here mainly the multiplicity dependent charged particle jet production in pp collisions at $\sqrt{s}$ = 13 TeV, based on the data collected by ALICE experiment during the LHC run in 2016. 

This analysis uses the minimum-bias (MB) events requiring at least one hit in either the V0 forward scintillators or the two innermost Silicon Pixel Detector layers (SPD) of the Inner Tracking System (ITS). The main detectors used for event selection and charged particle multiplicity categorization are the forward scintillator arrays (V0). The V0 detector consists of two scintillator detectors V0A and V0C,
 covering the pseudo-rapidity range of $2.8 < \eta < 5.1$ and $-3.7 < \eta < -1.7$, and they are used for triggering. The accepted events are required to have a primary vertex reconstructed within $|v_{z}| < 10{\rm cm}$ along the beam axis from the interaction point~\cite{ref:inclusivemult}.

Charged particles are reconstructed using combined information from the ITS and the Time Projection Chamber (TPC), which cover the full azimuth $0< \phi < 2\pi$ and $|\eta| < 0.9$ for charged hybrid tracks with $p_\mathrm{T,track} > 150 $MeV$/c$. Such charged tracks are then used to reconstruct charged particle jets via the anti-$k_\mathrm{T}$ algorithm from the FastJet package~\cite{ref:ktalgorithm}. The jet resolution parameters used in this analysis are $R$ = 0.2 and 0.4 with jet $p_\mathrm{T} > 0.1 $GeV$/c$. Reconstructed jets are further corrected for contributions from the underlying event to the jet momentum as: $p_\mathrm{T,ch~jet} = p_\mathrm{T,ch~jet}^{raw} - A_{\mathrm{ch~jet}}\cdot \rho_{\mathrm{ch}}$, where $p_\mathrm{T,ch~jet}^{raw}$ is the measured jet $p_\mathrm{T}$, $A_{\mathrm{ch~jet}}$ is the area of the jet and $\rho_{\mathrm{ch}}$ is the underlying background density obtained with the median occupancy method~\cite{ref:pbpb276shape}.

The main detector-related effects on the reconstructed charged jet spectrum are jet reconstruction efficiency, jet energy resolution (JER) and jet energy scale shift (JES). These effects are estimated by a jet response matrix, which is determined by a full detector simulation using Monte Carlo event generators. Such response matrices match the charged particle level jet $p_\mathrm{T}$ to the reconstructed jet $p_\mathrm{T}$ with different $\Delta R$ ($\Delta R = \sqrt{(\eta_\mathrm{{par}} - \eta_\mathrm{{det}})^2 + (\phi_{\mathrm{par}} - \phi_\mathrm{{det}})^2}$) conditions between the jet axes. MC event generators are also used to simulate the jet production to compare with data. 

After obtaining the jet response matrices, we use an unfolding procedure to account for the instrumental effects in measurements. In this analysis, the main algorithm for the unfolding of measurement jet $p_\mathrm{T}$ is the Singular Value Decomposition (SVD) method as implemented in the RooUnfold package ~\cite{ref:unfolding}. In addition, a Bayesian unfolding has been applied for validity checks and systematic comparisons. The unfolded spectra also need to be corrected for charged tracks reconstruction efficiency and track $p_\mathrm{T}$ resolution.

\section{Results}
 \subsection{Jet nuclear modification factor $R_{\mathrm{AA}}$}
\begin{figure*}[htbp]
 \begin{center}
 \includegraphics[width=0.48\textwidth]{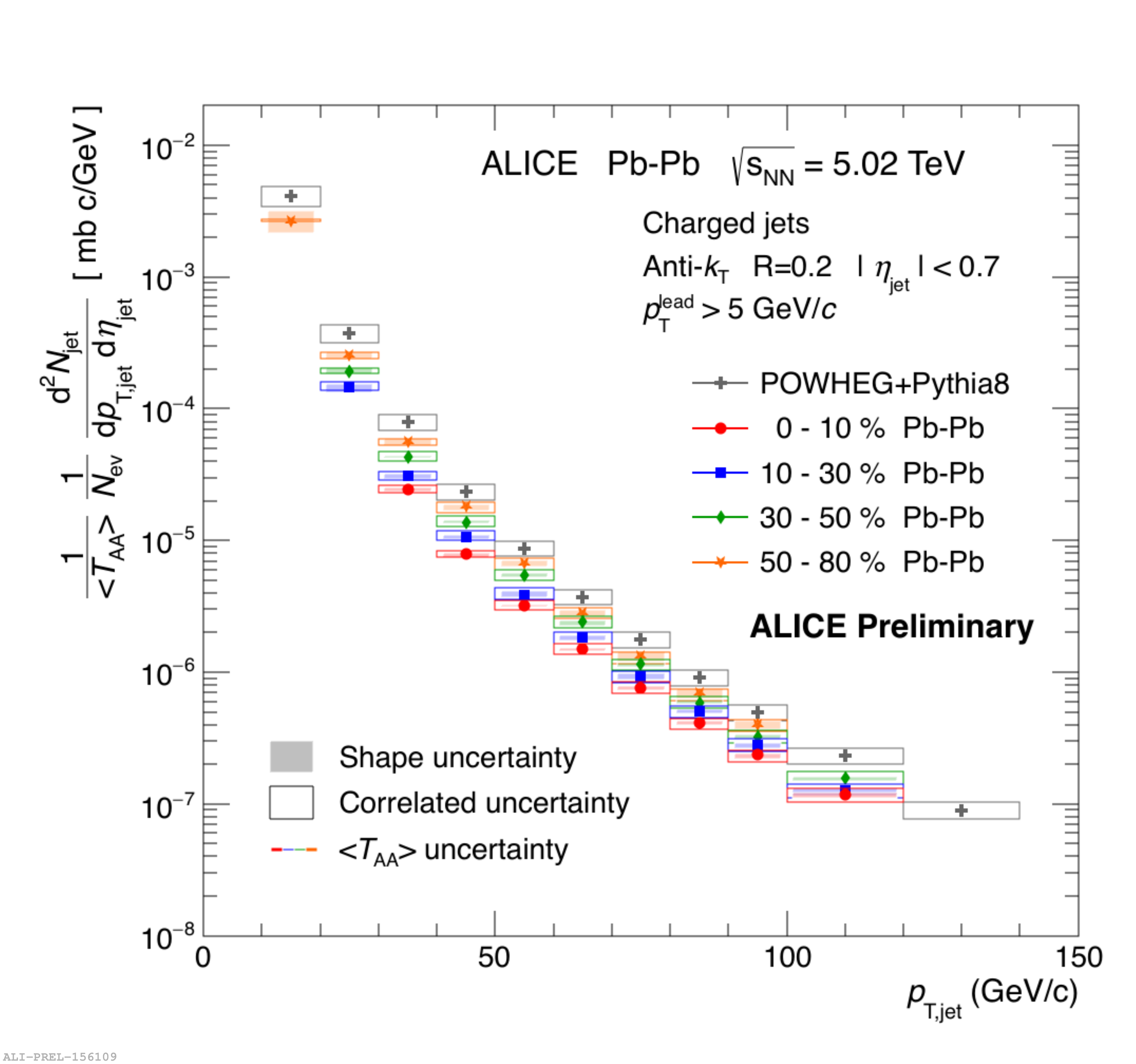}          
 \includegraphics[width=0.44\textwidth]{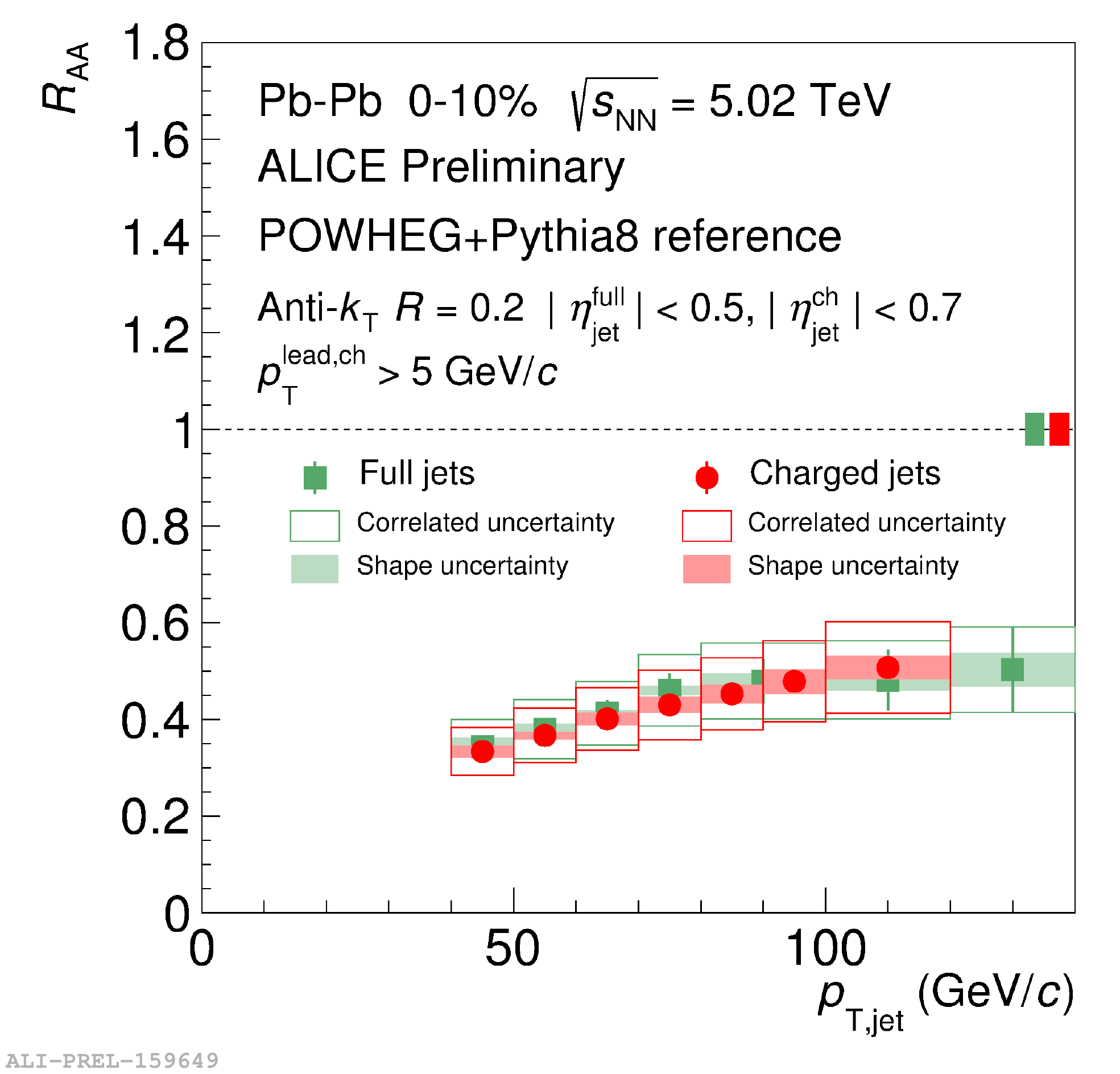}
 \end{center}
 \caption{Left: Jet yields for Pb-Pb collision with $R$ = 0.2,  Right: Nuclear modification factor $R_{AA}$ of full jets and charged jets}
 \label{fig:pbpbresults}
\end{figure*}

First, the unfolded jet spectra in different centrality bins with Pb-Pb collisions at $\sqrt{s_{\rm NN}}$ = 5.02 TeV for $R = 0.2$ are shown in Fig.~\ref{fig:pbpbresults} left, after $T_{\mathrm{AA}}$ scaling, where $T_{\mathrm{AA}} \equiv N_\mathrm{{AA}} / \sigma_{inel}^{\mathrm{NN}}$ is the ratio between number of binary nucleon-nucleon collisions and nucleon-nucleon cross section~\cite{ref:pbpb276shape}. To compare the spectra with jet measured spectrum in pp collisions, the jet nuclear modification factors $R_{\mathrm{AA}} = (\frac{d^{2}\sigma}{dp_{\mathrm{T}}d\eta}|_{\mathrm{AA}}) / (T_{AA}\frac{d^{2}\sigma}{dp_{\mathrm{T}}d\eta}|_{\mathrm{pp}})$ are obtained, which reflect the jet production suppression. Here we report the jet $R_{\mathrm{AA}}$ in more central nucleon-nucleon collisions in Fig.~\ref{fig:pbpbresults} right. These jet $R_{\mathrm{AA}}$ indicate strong suppression similar to charged particle jet $R_{\mathrm{AA}}$ and full jets result~\cite{ref:pppbpb276}. There is also a clear $p_\mathrm{T,jet}$ dependence, showing stronger suppression at lower $p_\mathrm{T,jet}$.

  \subsection{Inclusive jet cross section measurements}
\begin{figure*}[htbp]
 \begin{center}
 \includegraphics[width=0.44\textwidth]{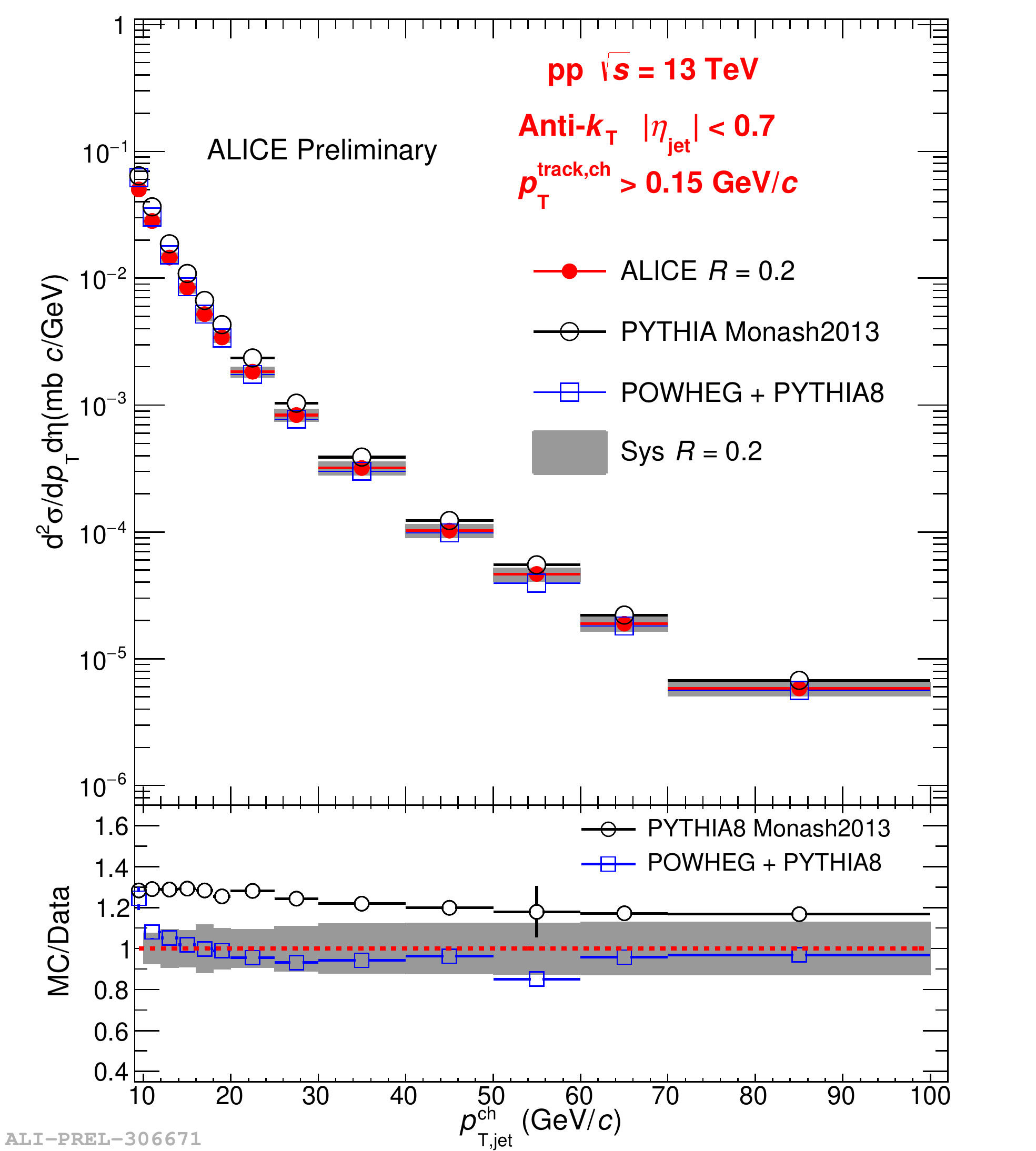}   
 \includegraphics[width=0.44\textwidth]{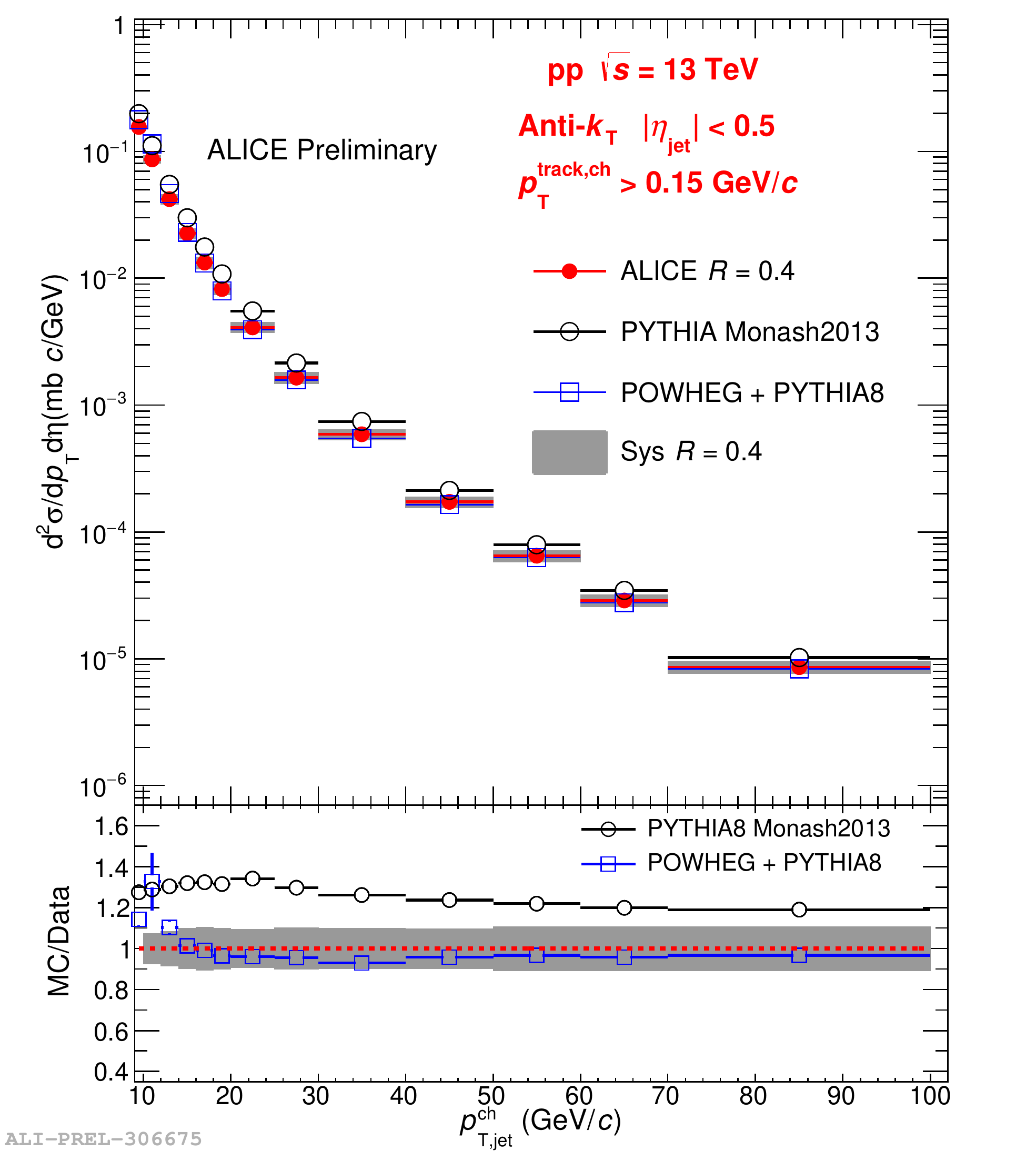}
 \end{center}
 \caption{ Comparison of inclusive jet cross sections to different MC simulations for $R$ = 0.2 (left) and 0.4(right)}
 \label{fig:ppinclusiveresults}
\end{figure*}

Second, Figure ~\ref{fig:ppinclusiveresults} shows the inclusive charged jet cross sections for $R = 0.2$ (left) and $R = 0.4$ (right) in pp collisions at $\sqrt{s}$ = 13 TeV using the $k_{\mathrm{T}}$ and anti-$k_{\mathrm{T}}$ jet finding algorithms. For jet cross sections of different jet resolution parameters $R$, the pseudo-rapidity ranges are limited to $|\eta| < 0.9 - R$, and transverse momenta from 10 to 100 GeV$/c$. The measurements are compared to predictions from the MC generators of LO PYTHIA8 Monash2013 and POWHEG, the latter of which is based on next-to-leading order (NLO) pQCD calculations~\cite{ref:pp7tev}. The ratios of MC simulations to ALICE data are shown in the bottom panels of Fig.~\ref{fig:ppinclusiveresults}. We can observe that POWHEG + PYTHIA8 can reproduce the data quite nicely at high $p_\mathrm{T,jet}$. Although the NLO calculation for inclusive charged jets description is closer to data, it still disagrees with the data in the low transverse momentum range~\cite{ref:pp502}. 

\begin{figure*}[htbp]
 \begin{center}   
 \includegraphics[width=0.88\textwidth]{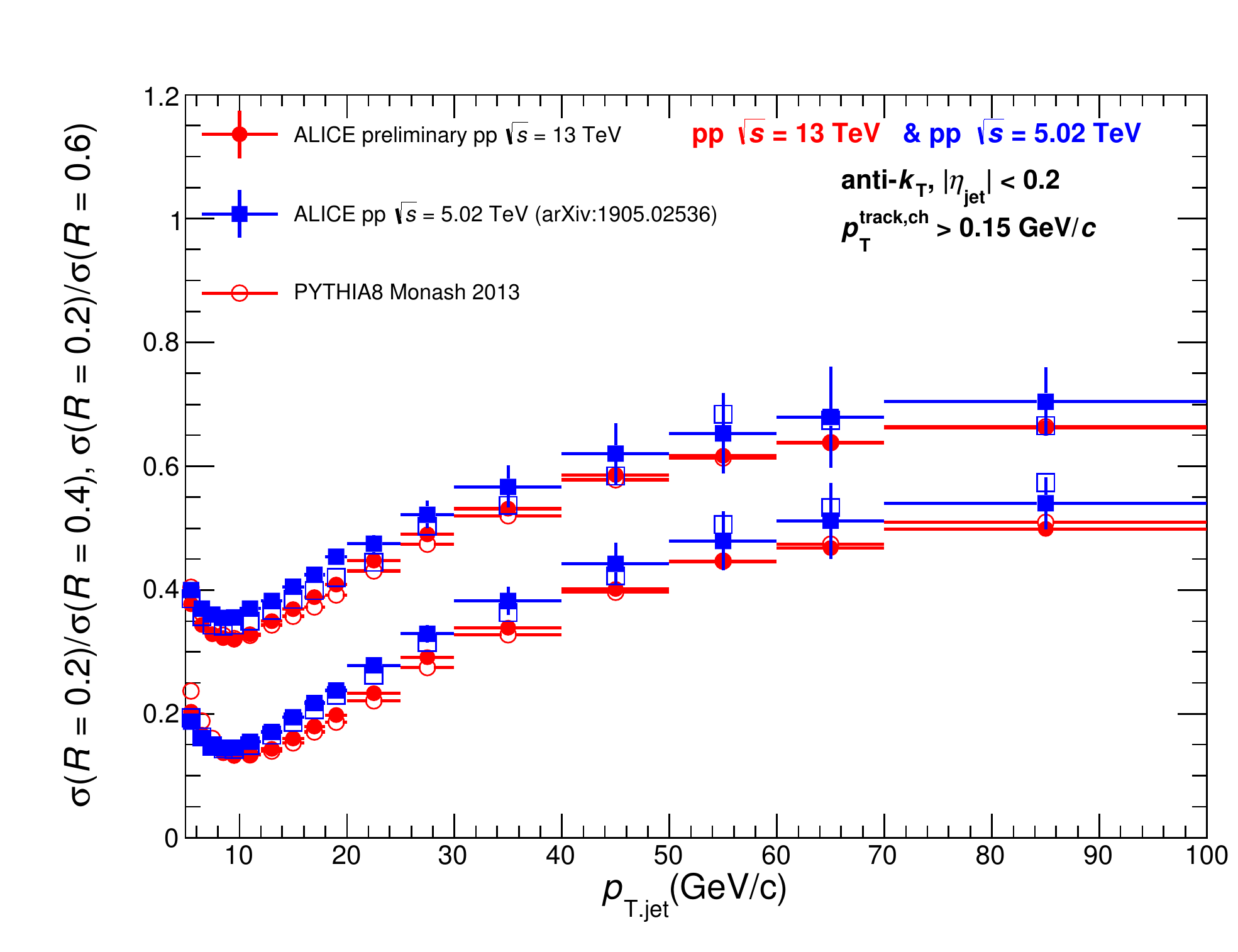}
 \end{center}
 \caption{ Charged jet cross section ratios for $\sigma(R = 0.2) / \sigma(R = 0.4)$ and $\sigma(R = 0.2) / \sigma(R = 0.6)$}
 \label{fig:ppinclusiveratio}
\end{figure*}

Next, we calculate the inclusive jet cross section ratios with resolution parameters $R = 0.2$ divided by $R = 0.4$ and $R = 0.6$ as shown in the Fig.~\ref{fig:ppinclusiveratio}, which increase with the jet transverse momentum. Since the cross section ratio measurements are sensitive to the collimation of particles around the jet axis, it indicates that jets are more collimated in the high $p_\mathrm{T,jet}$ region. The jet cross sections are measured in the same jet pseudo-rapidity range for the purpose of jet cross section ratio measurements, namely, within the range of $\eta < 0.3$, which is consistent with the fiducial jet acceptance for the largest resolution parameter studied ($R = 0.6$). The jet cross section ratio distributions are also compared with PYTHIA8 and POWHEG model calculations, which generally describe the data well within uncertainties.
The Figure ~\ref{fig:ppinclusiveratio} also presents a comparison of the jet cross section ratios between the ALICE Collaboration data in pp collisions at $\sqrt{s}$ = 13 TeV and $\sqrt{s}$ = 5.02 TeV~\cite{ref:pp502}. These two measurements show a similar increase with jet $p_\mathrm{T}$ and the jet is more collimated at higher $p_\mathrm{T,jet}$, while the ratio also has no strong dependence of the collision energies.  

 \subsection{Multiplicity dependent jet production}
\begin{figure*}[htbp]
 \begin{center} 
  \includegraphics[width=0.44\textwidth]{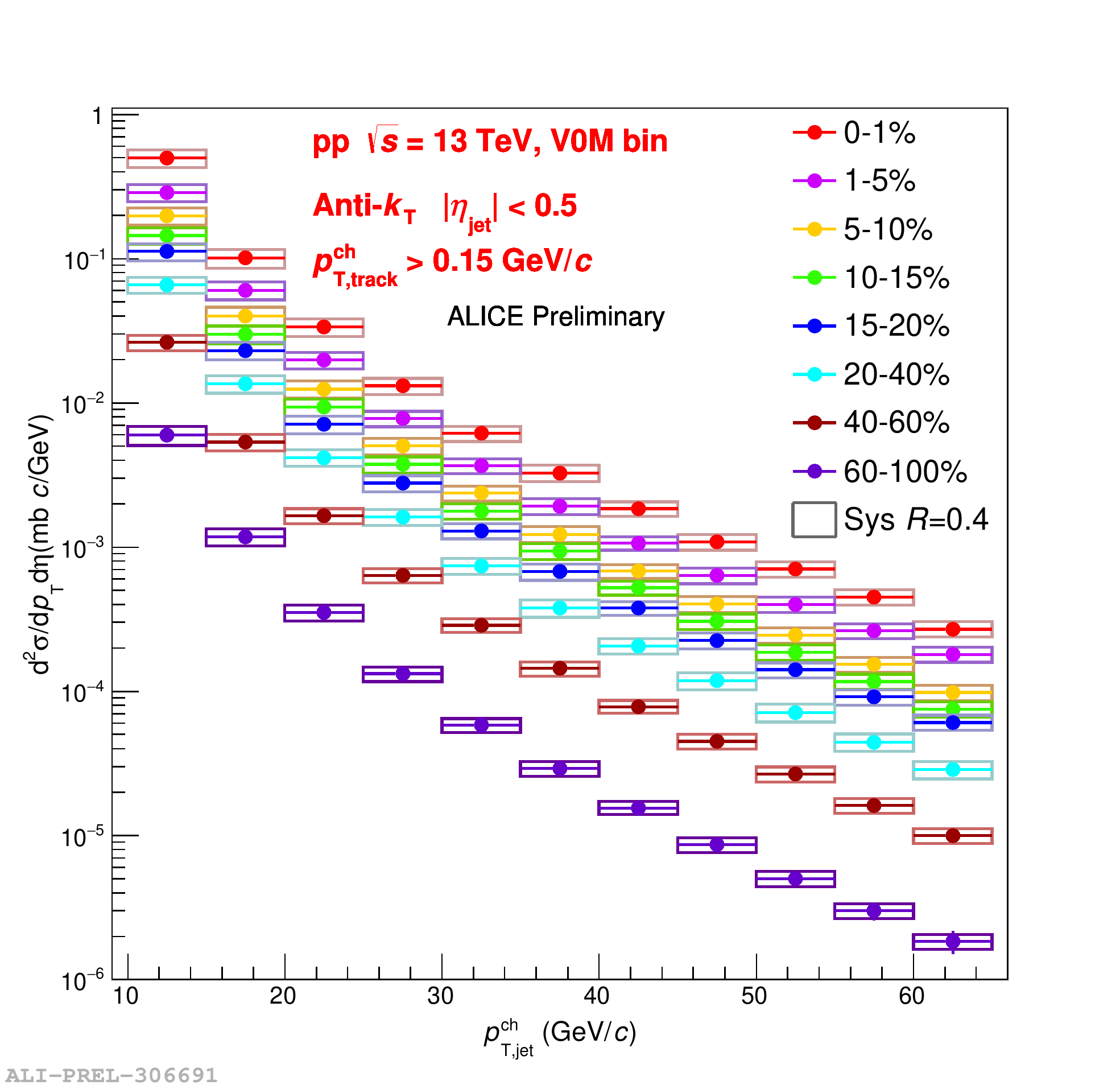}        
  \includegraphics[width=0.44\textwidth]{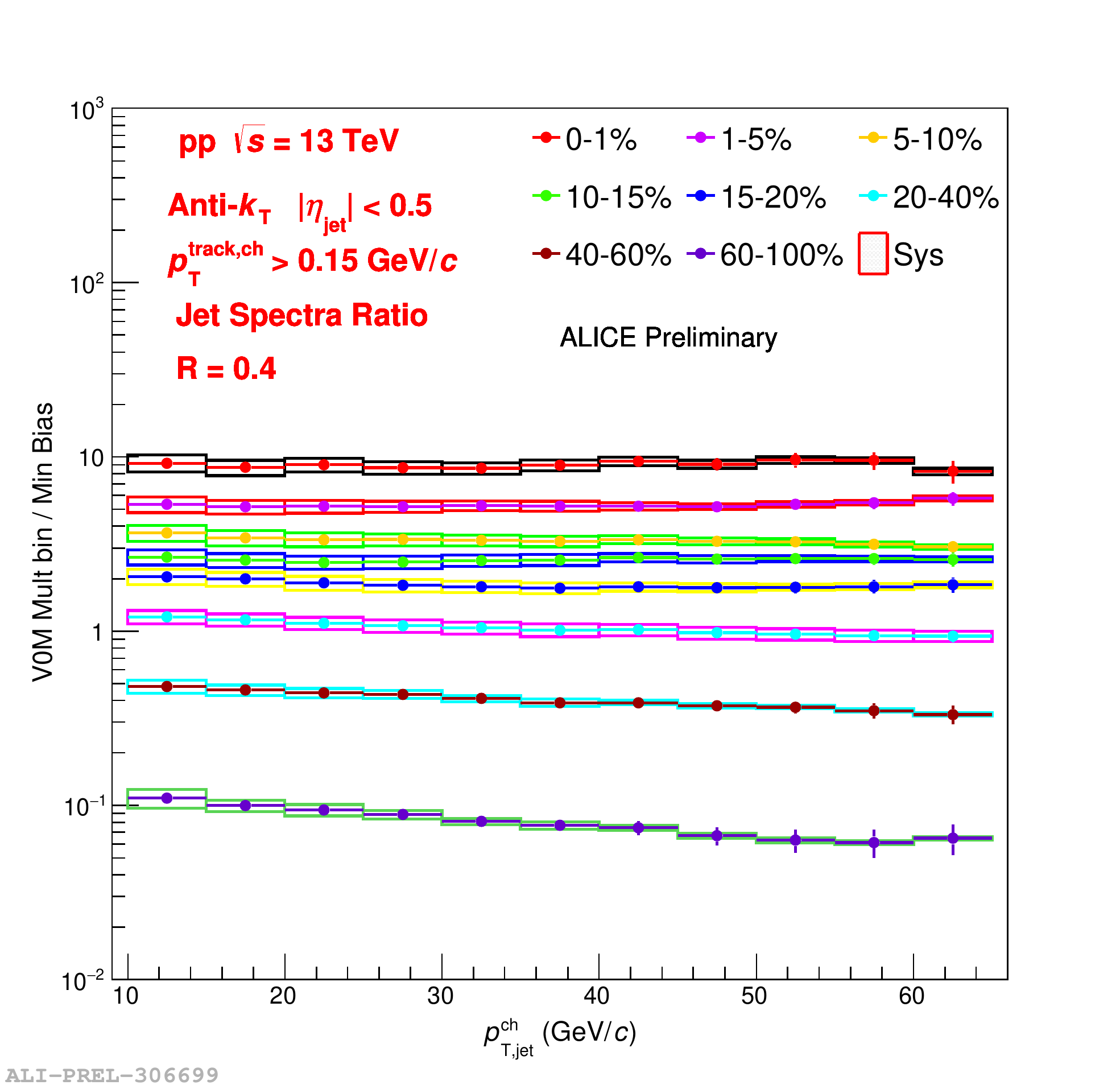}
 \end{center}
 \caption{Left: Jet cross sections for different multiplicity bins with $R$ = 0.4. Right: Jet production ratio between different multiplicity intervals to MB one}
 \label{fig:ppmultresults}
\end{figure*}

The corrected jet cross section for different multiplicity intervals using V0 estimator in pp collisions at $\sqrt{s}$ = 13 TeV are shown in Fig.~\ref{fig:ppmultresults} (left panel) as a function of jet $p_\mathrm{T}$ for $R = 0.4$. The jet cross sections have a clear increase from the lowest multiplicity bin to the highest multiplicity bin. To investigate the production modification in different multiplicity intervals, we calculated the jet production ratio in different multiplicity bins with respect to inclusive one (Fig.~\ref{fig:ppmultresults} right), namely the ratio of the multiplicity dependent jet spectra (Fig.~\ref{fig:ppmultresults} left) to the inclusive jet spectrum (Fig.~\ref{fig:ppinclusiveresults} right). From the ratio, we observed that in the highest multiplicity event (0-10\%), the charged jet production increases by a factor 10 compared to minimum-bias jet production. Whereas the cross section in the lowest multiplicity percentile ($60 - 100$\%) is only 10\% with respect to the inclusive one. The production ratio has weak $p_\mathrm{T}$ dependence, which indicated jet shape didn't change strongly between in the multiplicity intervals to minimum-bias one.

\begin{figure*}[htbp]
 \begin{center}
 \includegraphics[width=0.92\textwidth]{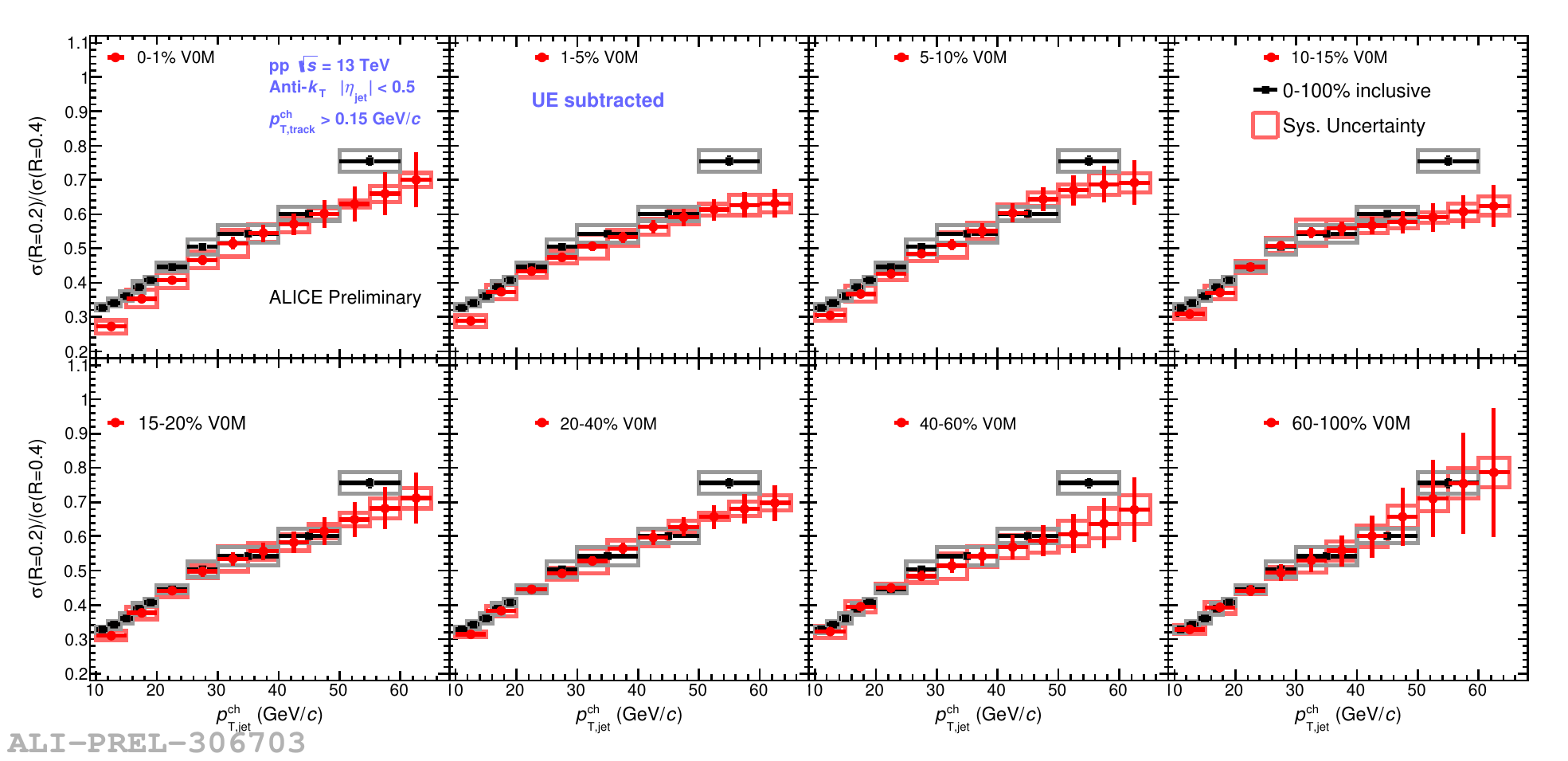}          
 \end{center}
 \caption{The charged jet cross section ratios of $R$ = 0.2 to $R$ = 0.4 for various multiplicity bins and MB one}
 \label{fig:8binsratio24}
\end{figure*}

Fig.~\ref{fig:8binsratio24} shows the ratios of jet production cross sections reconstructed with $R = 0.2$ and $R = 0.4$ for all multiplicity classes, which present the expected stronger jet collimation at higher jet $p_\mathrm{T}$. Furthermore, the ratios for different multiplicity events are all consistent with the result obtained in minimum-bias pp collisions~\cite{ref:ppb502}. This leads to the conclusion that jet cross section ratios using different jet resolution parameters have weak multiplicity dependence.

\section{Conclusions}
We report the transverse momentum $p_\mathrm{T,jet}$ spectra of jets reconstructed from charged particles in Pb-Pb collisions at $\sqrt{s_{\rm NN}}$ = 5.02 TeV with the ALICE detector at the LHC, as well as jet nuclear modification factor $R_{\mathrm{AA}}$ in more central (0-10\%) Pb-Pb collisions. The inclusive jet cross section in pp collisions at $\sqrt{s}$ = 13 TeV is measured for jet resolution parameters $R = 0.2$ and 0.4 with 10 GeV$/c < p_\mathrm{T,jet} < 100 \mathrm{GeV}/c$, and these results are compared to LO and NLO pQCD theoretical calculations. The NLO pQCD (POWHEG + PYTHIA8) calculation can reproduce the data better. 

Multiplicity dependences of charged jet production in pp collisions at $\sqrt{s}$ = 13 TeV have also been measured for transverse momentum from 10 GeV$/c$ to 100 GeV$/c$ with jet resolution parameters $R = 0.2$ and 0.4. The multiplicity classification is performed using forward detector V0. The jet productions in higher multiplicity events are larger than those in lower multiplicity events. Jet production ratios as a function of multiplicity, using charged jet cross sections in different multiplicity intervals with respect to minimum-bias one, has weak $p_\mathrm{T,jet}$ dependence in all multiplicity bins presented. The jet cross section ratio for $R = 0.2$ and 0.4 shows no strong multiplicity dependence, indicating no change of the degree of collimation of the jets between different multiplicity percentiles and inclusive one. These ratios have weak collision energy dependence.


\begin{thebibliography}{99}
%
\bibitem{ref:pp7tev} 
  B.~Abelev {\it et al.} [ALICE Collaboration],
  ``Charged jet cross sections and properties in proton-proton collisions at $\sqrt{s}=7$ TeV,''
  Phys.\ Rev.\ D {\bf 91}, no. 11, 112012 (2015)
\bibitem{ref:jetquenching}
  A.~Majumder and M.~Van Leeuwen,
  ``The Theory and Phenomenology of Perturbative QCD Based Jet Quenching,''
  Prog.\ Part.\ Nucl.\ Phys.\  {\bf 66}, 41 (2011)
  [arXiv:1002.2206 [hep-ph]].
\bibitem{ref:inclusivemult} 
  J.~Adam {\it et al.} [ALICE Collaboration],
  ``Measurement of charged jet production cross sections and nuclear modification in p-Pb collisions at $\sqrt{s_\mathrm{NN}}$ = 5.02 TeV,''
  Phys.\ Lett.\ B {\bf 749}, 68 (2015)
\bibitem{ref:multstrange} 
  J.~Adam {\it et al.} [ALICE Collaboration],
  ``Enhanced production of multi-strange hadrons in high-multiplicity proton-proton collisions,''
  Nature Phys.\  {\bf 13}, 535 (2017)
\bibitem{ref:ALICE} 
  B.~Abelev {\it et al.} [ALICE Collaboration],
  ``Performance of the ALICE Experiment at the CERN LHC,''
  Int.\ J.\ Mod.\ Phys.\ A {\bf 29}, 1430044 (2014)
  [arXiv:1402.4476 [nucl-ex]].
%
\bibitem{ref:ktalgorithm} 
  M.~Cacciari, G.~P.~Salam and G.~Soyez,
  ``The anti-$k_t$ jet clustering algorithm,''
  JHEP {\bf 0804}, 063 (2008)
\bibitem{ref:unfolding} 
  T.~Adye,
  ``Unfolding algorithms and tests using RooUnfold,''
  arXiv:1105.1160 [physics.data-an].
\bibitem{ref:pbpb276shape} 
  S.~Acharya {\it et al.} [ALICE Collaboration],
  ``Medium modification of the shape of small-radius jets in central Pb-Pb collisions at $\sqrt{s_{\mathrm {NN}}} = 2.76\,\rm{TeV}$,''
  JHEP {\bf 1810}, 139 (2018)
  [arXiv:1807.06854 [nucl-ex]].
\bibitem{ref:pppbpb276}
  S.~Acharya {\it et al.} [ALICE Collaboration],
  ``Measurements of inclusive jet spectra in pp and central Pb-Pb collisions at $\sqrt{s_{\rm{NN}}}$ = 5.02 TeV,''
  arXiv:1909.09718 [nucl-ex].
\bibitem{ref:pp502} 
  S.~Acharya {\it et al.} [ALICE Collaboration],
  ``Measurement of charged jet cross section in pp collisions at ${\sqrt{s}=5.02}$ TeV,''
  arXiv:1905.02536 [nucl-ex].
\bibitem{ref:ppb502} 
  J.~Adam {\it et al.} [ALICE Collaboration],
  ``Centrality dependence of charged jet production in p-Pb collisions at $\sqrt{s_\mathrm{NN}}$ = 5.02 TeV,''
  Eur.\ Phys.\ J.\ C {\bf 76}, no. 5, 271 (2016)
  [arXiv:1603.03402 [nucl-ex]].
\end{thebibliography}
\end{document}